\documentclass{vgtc}                          

\ifpdf
  \pdfoutput=1\relax                   
  \usepackage{graphicx}                
  \DeclareGraphicsExtensions{.pdf,.png,.jpg,.jpeg} 
\else
  \usepackage{graphicx}                
  \DeclareGraphicsExtensions{.eps}     
\fi%

\graphicspath{{figures/}{pictures/}{images/}{./}} 

\usepackage{microtype}                 
\PassOptionsToPackage{warn}{textcomp}  
\usepackage{textcomp}                  
\usepackage{mathptmx}                  
\usepackage{times}                     
\usepackage{cite}                      
\usepackage{tabu}                      
\usepackage{booktabs}                  

\PassOptionsToPackage{hyphens}{url}\usepackage{hyperref}

\onlineid{0}

\vgtccategory{Research}

\vgtcinsertpkg

\usepackage{subcaption}

\title{Towards a Theory of Bullshit Visualization}

\author{Michael Correll\thanks{e-mail: mcorrell@tableau.com}\\ %
        \scriptsize Tableau Research %
}

\abstract{%
In this unhinged rant, I lay out my suspicion that a lot of visualizations are bullshit: charts that do not have even the common decency to intentionally lie but are totally unconcerned about the state of the world or any practical utility. I suspect that bullshit charts take up a large fraction of the time and attention of actual visualization producers and consumers, and yet are seemingly absent from academic research into visualization design.
} 

\begin{document}


\firstsection{Introduction}

\maketitle

Our field heaps much of its scorn on two major categories of visualizations failures. The first are visualizations that are \textit{illegible} or otherwise \textit{hard to read}, such as the much maligned rainbow color map~\cite{borland2007rainbow} or 3D pie chart~\cite{o2018testing}. The second are visualizations that are purported to be \textit{deceptive}, for instance those that \textit{exaggerate} effect sizes (such as by truncating the y-axis of a bar chart to start from something zero~\cite{correll2020truncating}, or flip the y-axes such that lower values are higher up~\cite{pandey2015deceptive}). But these two categories are such a small part of what makes a visualization \textit{work} (or, more interestingly, \textit{fail to work}) as to be absurd: I can produce lots of things that are legible and not intentionally misleading and yet still fail to reflect genuine findings~\cite{mcnutt2020surfacing} or do anything \textit{useful}. To me what is potentially more dangerous than the handful of ``black hat'' visualization designers~\cite{correll2017black} is the vast amount of \textit{bullshit} charts we generate. For every intentionally misleading line graph created in earnest I have no doubt that there are several orders of magnitude more charts created that are just utter bullshit.

I use ``bullshit'' here in the technical sense, cribbed from Harry Frankfurt's \textit{On Bullshit}~\cite{bullshit}. Per Frankfurt, both an honest person and a liar are at least \textit{interested} in the truth. The honest person wants to tell you the truth, so they have to know what the truth \textit{is}, whereas the liar wants to convince you of something that they know is \textit{not} true, which also entails that they know what is \textit{actually} true and are just choosing to say the opposite. They are both at least playing the same game of having to engage with reality. The ``bullshit artist,'' by contrast, could not care less about what is true or false: their rhetorical goal is just to say whatever will accomplish their aim, and they don't much care if what they are saying is true or not. It is this disregard for reality that becomes, per Frankfurt, pernicious and corrupting. Graeber extends the notion of bullshit-related alienation to the increasing prevalence of ``bullshit jobs''~\cite{graeber2018bullshit}: work ``that is so completely pointless, unnecessary, or pernicious that even the employee cannot justice its existence''. As a result, bullshit jobs perform ``profound psychological violence''~\cite{graeber2013phenomenon} upon the people who work at them. Graeber's typology of ``bullshit jobs'' contains categories like \textit{flunkies}---``jobs that are just there to make someone else (generally immediate superior) look good or feel good about themselves'' and ``box-tickers''---jobs that ``are just there to make an organization feel it's doing something it really isn't,'' and occur in all sorts of work environments, even private sector employers that pride themselves on efficiency or productivity. 

It is my contention that many (most?) visualizations are bullshit in either the Frankfurt sense of attempting to persuade while being totally disconnected from notions of truths about the world, or in the Graeber sense of being created for no readily apparent purpose and to no real useful end. When they inform, these bullshit charts do so in the most superficial and unsatisfying ways--- for instance, a bar chart might tell me that so-and-so many widgets were sold during a particular period of time (\autoref{fig:bar}), but the CEO of the widget company will make widget-related business decisions based on some half-remembered self-help book called something like ``Widgeting Your Way To Wealth'' rather than by looking at the data, the widget salespeople are not foolish enough to blindly trust their widget-sales-tracking data no matter how nicely it is visualized on a PowerPoint slide, and you and I don't care about widgets anyway, except if we get \textit{really} desperate for anecdotes at a party (``did you read that article about widgets? as I recall, sales are going up''). So the bar chart of widget sales is mostly or entirely useless: it would not have mattered if the numbers in it reflected any sort of reality about widgets. And because of this uselessness, existing design principles about what makes a visualization ``good'' fail to apply either: I don't care about the data, so I don't care the ``data-ink ratio'' or presence or absence of ``chart junk'' either.

I focus on bullshit charts here because the limitation of bad charts to the deceptive and the illegible is an easy way for visualization designers and researchers to fall victim to what DeMarco calls ``The Fatal Premise''~\cite{myers2012responsible}: ``Evil is done by evil people; I am not an evil person and therefore I cannot do evil.'' If we assume that all bad visualizations are either the work of propagandists cackling as they try to convince you that no, widget sales are really going \textit{down}, or the work of hapless newcomers who just need to read a Tufte book or two to figure out that people would see those widget sales much more clearly if they took out the axis gridlines from their bar chart, then we feel no need to question our field or ourselves (maybe we'll throw a sop to ``shucks, if only people had more data literacy'' or ``shucks, if only our charting libraries had better defaults''). After all, \textit{we} (and all of our friends) aren't evil or hapless! But we have all made bullshit charts, with bullshit data, for bullshit ends, and so we have no excuses to run from our responsibilities here. 

To illustrate the pervasiveness of bullshit charts, I will spend much of this paper enumerating some of examples of bullshit charts, vaguely gesture towards a taxonomy or other categorization schema for bullshit charts, and end with a perfunctory call to do something about all the bullshit.

\section{This Chart Could Have Been An Email}

\begin{figure}
    \centering
    \includegraphics[width=0.35\columnwidth]{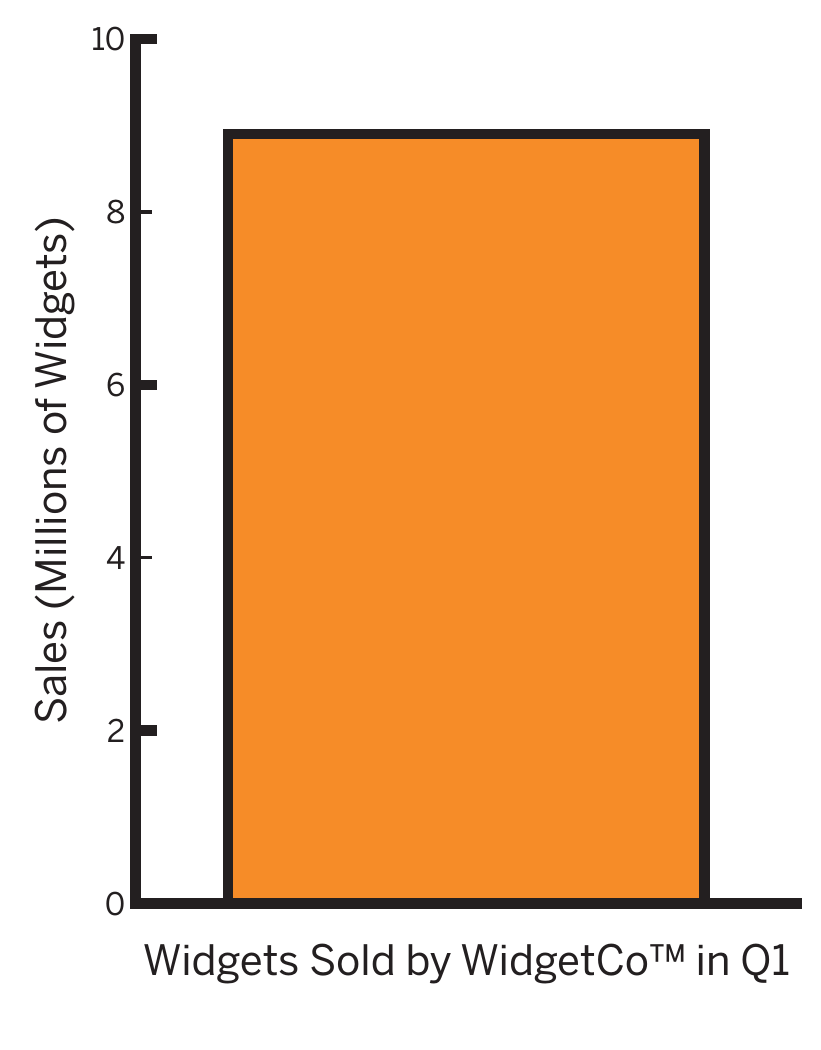}
    \caption{An example of ``number decoration.'' This chart could be entirely replaced by a sentence on the order of ``WidgetCo sold approximately 9 million widgets in Q1'' and work just as well--- the bar chart adds nothing. In fact, since WidgetCo does not exist, and I pulled that number out of thin air, this chart could fail to exist \textit{entirely} and I wouldn't lose any sleep over it.}
    \label{fig:bar}
\end{figure}

In describing the potential pitfalls in industry-led data visualization, Stefaner~\cite{moritz} refers to the phenomena of ``number decoration'':
\begin{quote}
Simple KPIs (key performance indicators) are decorated with visual elements to look sexy in dashboard or reports, but you never get access to the full texture and variety of the data underlying those numbers. It’s like you were interested in the forest, but you only get a beautiful little stick. This is not harnessing the power of visualization--- this is dressing up numbers. 
\end{quote}

Stefaner was referring specifically here to dashboards that augment their KPIs with unneccessary little donut charts or bars with arrows or what have you, but I think there is a wide range of scenarios where a visualization could have been just a table, or a just a number, or just a sentence, or even failed to exist entirely, without causing anybody any grief (see \autoref{fig:bar}). While I like the concept, I must admit ``decoration'' is a bit of a red herring: I am not referring here to ``chart junk'' in the Tufte-ist sense, of annotations to graphs that do not strictly encode data--- a picture of a happy bird next to my chart might at least do something useful like remind me that my data is about birds~\cite{borkin2013makes} or draw the bird-liking audience to my chart in a sea of birdless alternatives: decorate your chart all you want, I don't care. Rather, my focus here is on charts whose sole purpose seems to be to lend an air of authority to the data, to make it look ``science-y'' or ``statistics-y'' because it can't be totally nonsense if there's a bar chart next to things, right? Although n.b. that it's unclear if these sorts of attempts to use charts for these sorts of ethical appeals even work~\cite{dragicevic2017blinded,richards2003argument}. My rule of thumb here is that if your visualization could be replaced by, say, a stock photo of an arrow going up (or down) without changing the intended message or use of what you've built, then it's probably a bullshit visualization, or at the very least a tendril in the ever-expanding octopus of what I term ``Potemkin Data Science''~\cite{correll2020}--- data-related work that is performed for largely no real purpose and to no useful effect.

This mismatch between the stated goals of visualization (communicating nuanced and important information to wide audiences) and the goal of number decoration (to make some numbers look pretty) infects our abstractions around visualization as an abstract discipline. For instance, our conception of what makes a ``good'' visualization is often limited by thinking about the efficient extraction of individual values~\cite{bertiniscatter}, producing a kneejerk academic reaction against charts that encourage deeper thought, engagement, or analysis.

A related genre of valueless visualization is the more pernicious notion of what Froehlich and I have called the ``spectacular dashboard''~\cite{correlldash}: a dashboard that exists mostly as a ``spectacle'' or (occasionally more nefariously) as ``decision-laundering'': to convince you that, yes there is all of this big and complex data \textit{somewhere}, and yes, you can maybe see interesting trends, but where you, the viewer, have no apparent way to actually \textit{do anything} with the numbers you are looking at: you are just supposed to \textit{look} at the big complicated dashboard, maybe root for the ``home team'' (``I hope widget sales go up next year too!'') and be soothed or numbed or otherwise dissuaded from deviating from a status quo set by others. It's designed to make you think \textit{less}, not more.

\section{Let a Hundred Sharpiegates Bloom}

\begin{figure}
     \centering
     \begin{subfigure}[b]{0.85\columnwidth}
         \centering
         \includegraphics[width=\textwidth]{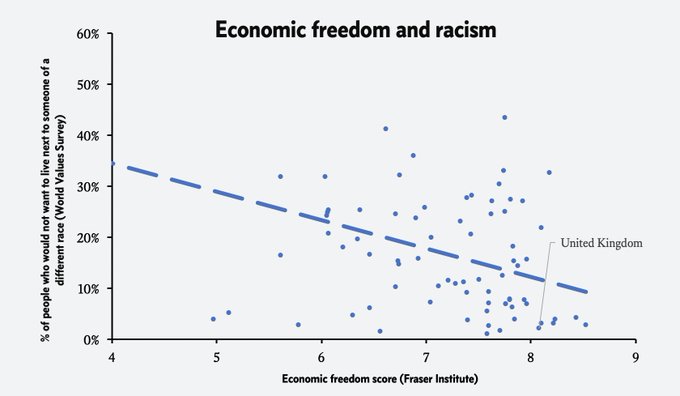}
         \caption{A graph used to claim that ``Countries with more economic freedom have less racist attitudes''~\cite{scatterplot}.}
         \label{fig:scatter1}
     \end{subfigure}
     
     \begin{subfigure}[b]{0.85\columnwidth}
         \centering
         \includegraphics[width=\textwidth]{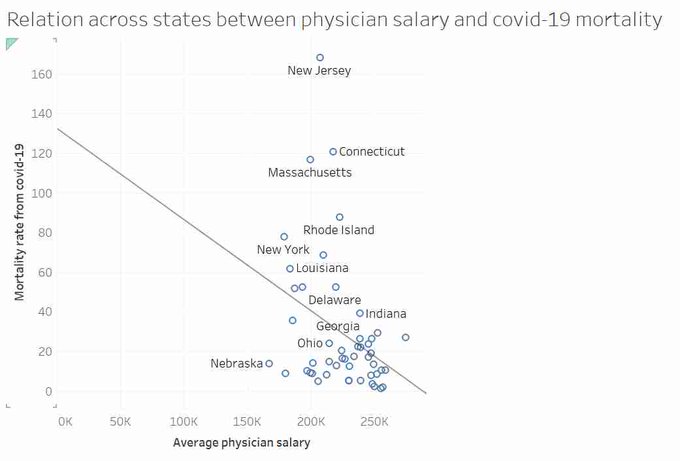}
         \caption{A graph used to claim that ``States where physicians are highly paid have lower COVID-19 mortality per capita''~\cite{scatterplot2}.}
         \label{fig:scatter2}
     \end{subfigure}
        \caption{Two entries in the emerging genre of ``plot some points, draw an arbitrary line, and then say whatever you want.'' Both authors of these charts were challenged about the poor fits in their data and the lack of an obvious strong correlation. The first responded ``Yeah it's $r^2 = 0.14366$. So clearly lots more going on across societies. But the point stands...'', while the second said ``The ubiquitous misuse and tyranny of SST [statistical significance testing] threatens scientific discoveries and may even impede scientific progress.'' My response to both is ``lol.''}
        \label{fig:scatterplots}
\end{figure}

\begin{figure}
    \centering
    \includegraphics[width=0.75\columnwidth]{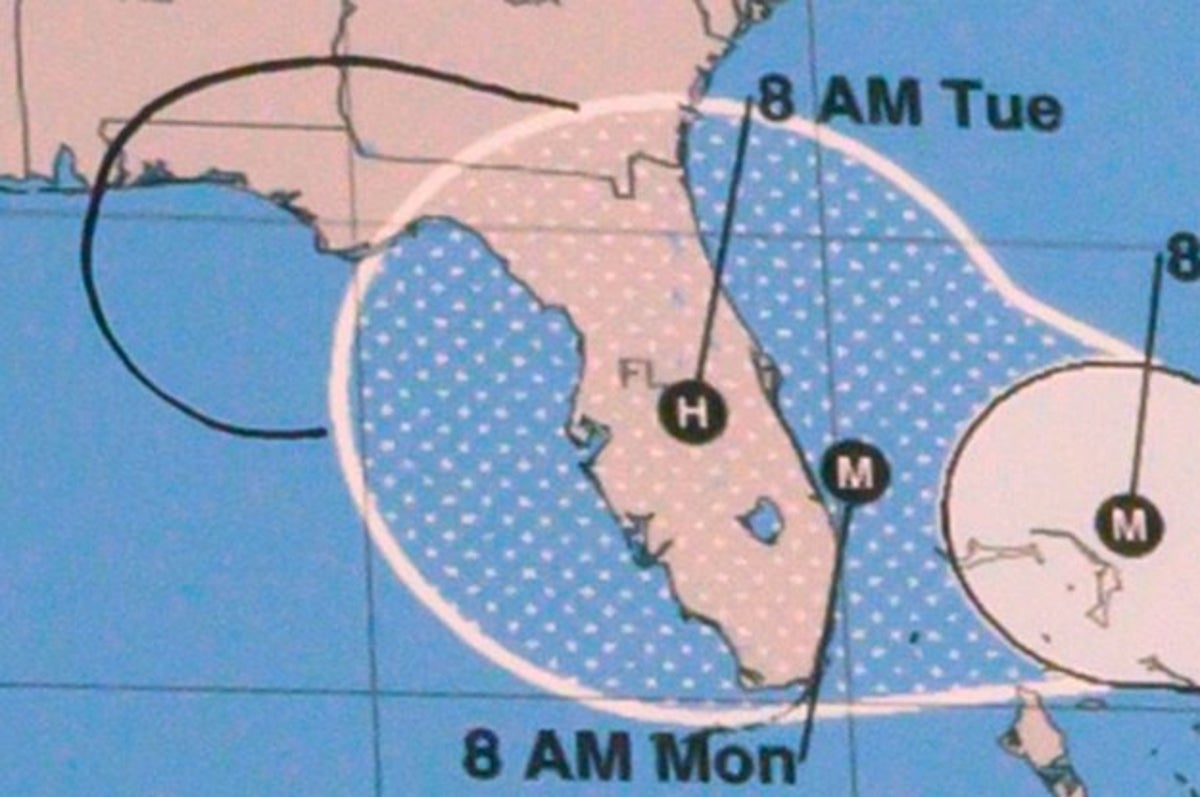}
    \caption{An inset of the inciting chart in ``Sharpiegate''~\cite{sharpiegate}, where a map of hurricane predictions was altered with a marker to include potential impact on Alabama in order to align with then-president Trump's claim that Alabama was likely to be impacted by Hurricane Dorian. Other parts of this affair were probably more serious (like allegedly threatening the senior staff of NOAA with termination if they didn't release an unsigned statement backing up the president), but this amateurish visual manipulation was the part that was the funniest, in a fatalist mirthless chuckle sort of way.}
    \label{fig:sharpiegate}
\end{figure}

Meeks~\cite{meeks2019} points to events like Sharpiegate (\autoref{fig:sharpiegate}) as representing part of a turning point in the modern history of data visualization: ``the United States has never had a president that cared more about the appearance of data than the data itself, until now.'' Caring about how the data \textit{look} in a chart, rather than the underlying \textit{truth or falsity} of the data, is about as straightforward of a definition of Frankfurt-style bullshit as you can ask for. People are starting to cotton on to the idea that charts have rhetorical force in and of themselves: they are showing ``the data,'' and it is so tempting to conflate ``the data'' with ``the truth.'' And so charts are taken more seriously than anecdotes or other claims that we investigate with more scrutiny, and people perform less of the epistemic hygiene and due diligence that they might do with, say, a written statement (I think a lot about the participants in a survey by Peck et al.~\cite{peck2019data} who were reluctant to consider the source of a chart when assessing its trustworthiness, connected with quotes like ``I think that information is information no matter from where it comes from''). In the terms of the Drucker~\cite{drucker2012humanistic} quote that I seem to be unable to avoid using in about half the things I write these days, when people look at a chart it is ``as if all critical thought had been precipitously and completely jettisoned''--- the chart is a picture of the world as it is, and the dataset within it unalterable truth. 

Given the pride of place that charts have (or, again, are perceived as having; it's not clear to me empirically that people seem to put undue weight on visualizations \textit{per se} over other ways of communicating data), there is now an emerging genre of dumb chart where you just plot data that look random, throw a trend line or confidence interval on top of all the mess, and declare victory for whatever point you want to make (\autoref{fig:scatterplots}).

There is so much complexity you can hide in a chart (the data collection process, the uncertainty, the model assumptions, the outliers--- whatever is inconvenient), and it's so much easier to share a jpg than an jupyter notebook or what have you, that just using an arbitrary chart as a backdrop to make an arbitrary claim and hoping that nobody checks too closely is a temptation to which I'm surprised more bullshit artists haven't succumbed.

I would like the reader to recall the fatal premise. I chose examples of bullshit like Sharpiegate because the manipulation is obvious (and a bit humorous). But we don't get an excuse for ignoring this issue just because we aren't directly mentioned in the rogue's gallery. Do we check to make sure that the conclusions from our visualization are sensitive to different choices of methods~\cite{dragicevic2019increasing}? Do we think we're free from all of the self-serving bits of the replication crisis~\cite{kosara2018skipping}, like the file drawer problem (we publish the visualizations that support a belief but ignore the ones that don't) or the multiple comparisons problem~\cite{zgraggen2018investigating} (we've made so many different graphs that at least some of them, by chance alone, have a pattern we think is interesting), or any of the whole host of visualization ``mirages''~\cite{mcnutt2020surfacing} that mean that what you see is not what you get? Our sharpies are a little bit more subtle, but if we think we don't sometimes use visualizations to fool ourselves, we're fooling ourselves.

\section{The Data-Ink Ratio is the First Casualty of War}

\begin{figure}
    \centering
    \includegraphics[width=\columnwidth]{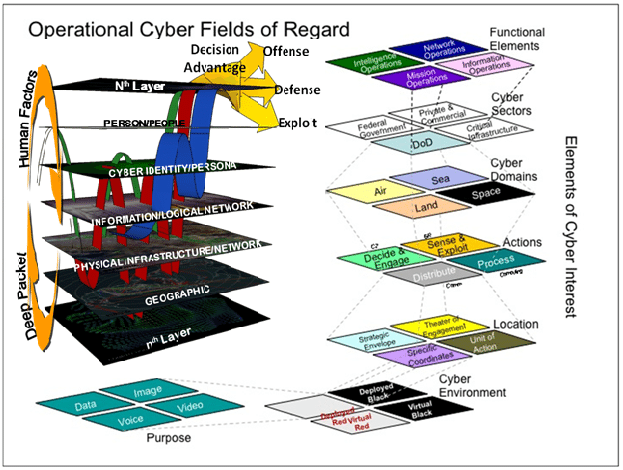}
    \caption{An example from Hwang's~\cite{defense} @DefenseCharts Twitter account. No idea of the provenance on this one, but also no idea about many things that have to do with this chart. The word ``cyber'' only appears five times by my count, which is on the low end for this particular genre.}
    \label{fig:dod}
\end{figure}


No description of bullshit charts would be complete without a giving pride of place to the efforts of the U.S. military-industrial complex. Hwang's~\cite{defense} @DefenseCharts Twitter account, ``dedicated to the presentational aesthetics of the defense-industrial complex,'' collects some of the more interesting examples: flow charts filled with seemingly arbitrary nouns (the word ``Cyber'' usually appears at least once, see \autoref{fig:dod}) and sharing space with color gradients and 3D shapes and clip art of soldiers or vehicles to contribute to an instantly recognizable and totally baffling style of information design. These charts serve a purpose, but you would have to be charitable in the extreme to claim that this purpose was primarily to inform. Indeed, the ``Afghanistan Papers''~\cite{afghanistanpapers}, the cache of documents connected to the American invasion of Afghanistan, shows that bullshit charts (and their close cousin, the bullshit slide deck~\cite{powerpoint}) were at the heart of the effort to convince external audiences that the war was going well while the realities on the ground were quite different:
\begin{quote}
John Garofano, a Naval War College strategist who advised Marines in Helmand province in 2011, said military officials in the field devoted an inordinate amount of resources to churning out color-coded charts that heralded positive results.

``They had a really expensive machine that would print the really large pieces of paper like in a print shop,” he told government interviewers. ``There would be a caveat that these are not actually scientific figures, or this is not a scientific process behind this.''

But Garofano said nobody dared to question whether the charts and numbers were credible or meaningful.

``There was not a willingness to answer questions such as, what is the meaning of this number of schools that you have built? How has that progressed you towards your goal?'' he said. ``How do you show this as evidence of success and not just evidence of effort or evidence of just doing a good thing?''
\end{quote}

These bullshit visualizations, since they impart no clear information, ``relieve the briefer of the need to polish writing to convey an analytic, persuasive point''~\cite{bullshit}, and have been used to dilute responsibility or agency for decision-making. For instance, if I issue you a direct order and it ends poorly, I could conceivably be to blame. But if I give you a vague and complex slide deck instead of explicit instructions (as U.S. leaders were accused of doing~\cite{bullshit}), then I can always blame you for not understanding me, or for acting against my intentions.

In short, this brand of bullshit chart serves as a way to distract people from bad news with irrelevant metrics, bore the audience into complacency or indifference, or convince the audience of the presenter's intelligence or depth or knowledge (while at the same time removing the burden of making explicit falsifiable or actionable statements) through sheer overwhelming visual complexity. Bringing in a visualization designer to ``fix'' these designs would arguably make them \textit{less} effective for these purposes, because their purpose is \textit{not} to communicate clearly.

\section{Discussion}

\begin{figure}
    \centering
    \includegraphics[width=0.95\columnwidth]{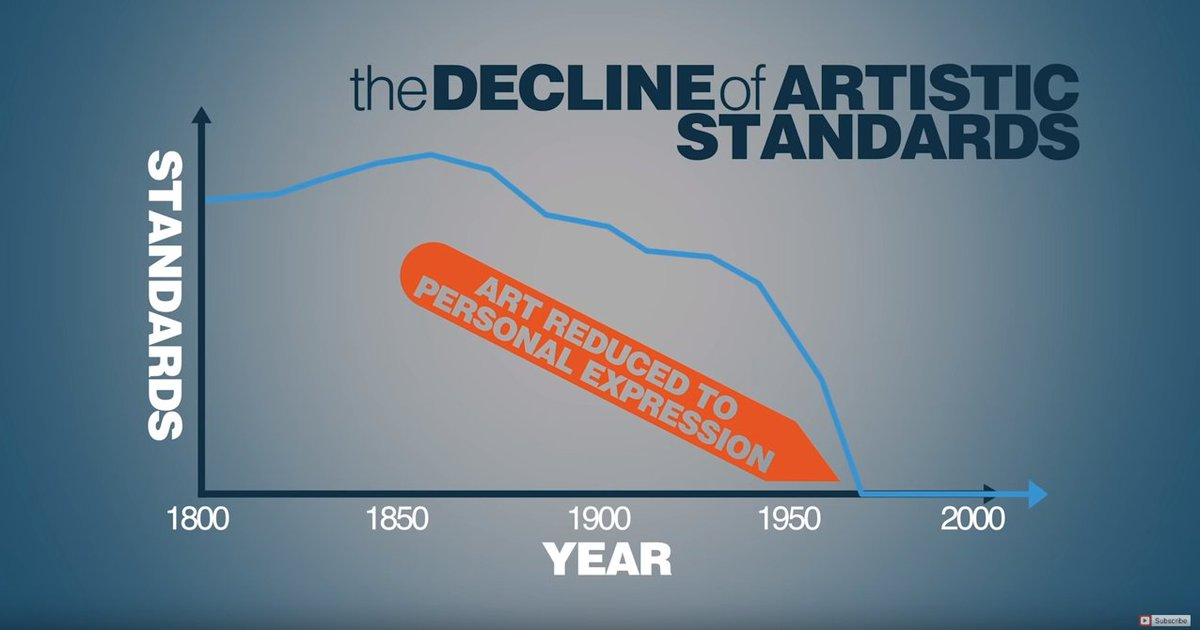}
    \caption{A graph used in a video~\cite{prager} by American conservative media company Prager U showing the nefarious impact of modern art. This graph uses a noisy line to make it look like there is a complex metric being measured precisely over time, when in fact there is not (and even if there \textit{were} somebody claiming to have an ``objective'' measure of artistic standards, would you trust them?). It is a chart largely uninterested in underlying reality at the expense of looking ``data-y'' and authoritative for a predetermined argument (viz., that modern and contemporary art is bad).}
    \label{fig:prager}
    \vspace{-1em}
\end{figure}

Time, space, and reader patience limited me to a handful of examples, but I hope you get the general idea: there is a class of visualization failures I call \textbf{bullshit visualizations}. These are visualizations that either, as per Frankfurt~\cite{bullshit}, are unconcerned with the actual truth or falsity of the world or, as per Graeber~\cite{graeber2018bullshit}, are generated for no real purpose other than keeping up appearances. Whenever you make a chart of whatever data you have on hand without checking to see if it is relevant, useful, or accurate for the goals you have in mind, a bullshit visualization is sure to follow. Bullshit visualizations are all too often a symptom of bullshit existing elsewhere (in society, in an organization, in the individual), and so fighting them often means fighting institutional battles at the scale beyond recommending a bar chart instead of a pie chart. And bullshit does not tend to limit itself to just a few bad eggs at the ends of particular bell curves, but acts as an insidious or corrupting force even for the well-intentioned.

I am reluctant to present a full taxonomy of types of visualization bullshit for several reasons. Firstly, I think it's a problem that awaits further empiricism (although I think efforts like Cairo's ``Visual Trumpery'' lecture series~\cite{cairotrump} are reasonable attempts to convince people of the scale and extent of certain kinds of chart bullshit). Second, once again I feel the fatal premise is breathing down our neck. A taxonomy is a temptation to say ``a ha, \textit{my} chart isn't on this list, so it must not be bullshit'' or alternatively ``this chart has one of the qualities on this list, so it must be bullshit and so can be entirely ignored.'' So, with those caveats in mind, here are a few patterns that are at least strongly correlated with bullshit:

\textbf{Decorative Chart:} Again, I am less concerned with the decorative \textit{elements} in charts that Tufte seems to care so much about, but the scenario where the charts \textit{themselves} are decoration. You've got to have charts in a paper or a slideshow in order to be taken seriously. ``Lack of colorful figures'' was a strongly diagnostic characteristic of a paper likely to be rejected in a (granted, somewhat tongue-in-cheek) attempt to model ``good'' and ``bad'' CVPR papers~\cite{von2010paper}. If there's no (or not enough) data to make a useful chart, you just make one that looks nice and put it in anyway.

\textbf{Stock Footage Chart:} A chart that exists mostly so you can have something on in the background to set the scene. It just needs to be vaguely related to the matter at hand. An example is all of the computer science talks that begin with a chart showing the exponential growth of data or computer users or processing power or some other vaguely computer science-y topic that serves only to set the scene for a topic that is really not about exponential growth at all. But you need a chart for visual interest at the beginning, otherwise your audience won't see another one for another few dozen slides and might decide to check their phones during your talk.

\textbf{Novocaine Chart:} If the stock footage chart is supposed to perk you up in an otherwise boring talk, Novocaine charts are the opposite. As with the previously discussed ``spectacular dashboards,'' these charts are supposed to numb you to the scale and complexity of the data. It is supposed to make the presenter look smart (after all, look at how complex of a chart they had to generate in order to present the data!) and/or the audience feel stupid or powerless (there's no way \textit{we} could do anything about this problem). This appeal to complexity and nuance is also meant to diffuse any need to make an actual decision or state an actual opinion.

\textbf{Texas Sharpshooter Chart:} When you draw your ``bullseye'' chart after you've already decided what your conclusion is going to be. If the resulting chart manages somehow to surprise or contradict you, it's the data that are wrong, and so you start looking for other tools in your toolkit to avoid being surprised. If the data don't support your conclusion, then you just get different data, or take out your sharpie, or otherwise minimize or alter the information in your dataset. While a lot of overtly deceptive techniques can be employed here, there are lots of other data habits that we generally assume are laudatory (like ``building skepticism'' or ``adding nuance'') that can be a cover for this sort of behavior.

Charts can and do cross these boundaries: for instance, \autoref{fig:prager} is a chart that decorates a small amount of data (in this case absolutely no data whatsoever) in an attempt to make it seem more ``data-y,'' does so in support of a pre-selected empirical conclusion, as part of a brand exercise designed to make it fit in. One thing it \textit{doesn't} do is clearly and accurately present data about the real world, but then again that wasn't really the point, was it?

I have conceptual and pragmatic concerns about anything like an automatic bullshit detection, but in the meantime here's rule of thumb vaguely connected with a notion of ``surprise'' (in the Bayesian sense~\cite{correll2016surprise}, but also the notion of Algebraic Visualization Design~\cite{kindlmann2014algebraic}, or graphical inference sense~\cite{wickham2010graphical}). My untested visualization first-pass approximation visualization bullshit detector is: \textbf{if the phenomena behind the data in your chart were completely different from what they are now, would your audience notice? And if they did notice, would they care?} If your weather chart performs about the same persuasive work in a world where hot snow falls up, then you might be in trouble.

I will inject a little bit of nuance here at the end and say that just because a visualization fails my test, or falls into one of my buckets of bullshit above (sorry for the imagery here) doesn't mean that it is necessarily bullshit, and certainly doesn't mean it is necessarily ``bad''. I can make a chart out of idle interest, artistic drive, pedagogical intent, or just as a test of my chart-making skills without being too worked up about whether the resulting chart is truthful or useful--- it think it's something about a shared communicative act where the bullshit happens. And a chart can of course \textit{show} us nothing (for instance, present a negative statistical result) without \textit{telling us} nothing (negative results are results too!): it's about being \textit{unconcerned} with reality rather than reflecting data that are noisier or less interesting than we had hoped.

I would like to end by asking how academic visualization practices contributes to bullshit (either through the charts or systems we generate, the techniques we champion, or the data culture we build), and ask us to stop. To put it as bluntly as possible, as academics we have certain perverse incentives: to publish or perish, to develop and promote novel techniques, to be perceived as doing complex and sophisticated work, and all sorts of bullshit. The charts we design for others to use, create to communicate our own results, or use as examples in our teaching, all have the potential to be bullshit, and in fact exist within structures that push them towards being bullshit. The fight against bullshit should begin with self-critique.

\bibliographystyle{abbrv-doi}

\bibliography{template}
\end{document}